The characteristics of a low background germanium gamma ray spectrometer at China JinPing underground Laboratory

Yuhao Mi[a,b], Hao Ma[a,b], Zhi Zeng[a,b*], Jianping Cheng[a,b], Jian Su[a,b], Qian Yue[a,b]

[a] Department of Engineering Physics, Tsinghua University, Beijing 100084, China

[b] Key Laboratory of Particle and Radiation Imaging (Tsinghua University), Ministry of Education, Beijing 100084, China

Abstract: A low background germanium gamma ray spectrometer, GeTHU, has been installed at China JinPing underground Laboratory. The integral background count rate between 40 and 2700 keV was 0.6 cpm, and the origin was studied by Monte Carlo simulation. Detection limits and efficiencies were calculated for selected gamma peaks. Boric acid and silica sand samples were measured and $^{137}$Cs contamination was found in boric acid. GeTHU will be mainly used to measure environmental samples and screen materials in dark matter experiments.

1. Introduction

Low level high purity germanium (HPGe) gamma ray spectrometry has been developed for many years and applied in different fields such as fundamental physics researches and conventional sample investigations (Hult, 2007). Recently, as detections of rare events such as neutrino interaction, dark matter and neutrinoless double beta decay have accelerated and monitoring of environmental radioactivity has been highly focused on, low level HPGe gamma spectrometers are playing an increasingly important role in material selections for rare event experiments and measurements of environmental samples (Arpesella et al., 2002; Budjáš et al., 2009; Kohler et al., 2009).

Generally, backgrounds of low level HPGe spectrometers have three origins: a) cosmic rays, including backgrounds directly from cosmic rays and indirectly from secondary radiations; b) environmental radioactivity, mainly referring to the natural radioactive series, that is, the $^{238}$U and $^{232}$Th series, and the single radionuclide, $^{40}$K, and some neutron sources will be taken into consideration as well, such as spontaneous fissions of heavy nuclides and (α,n) reactions; c) radioactivity in construction materials and detectors (Heusser, 1995; Hult et al., 2006; Laubenstein et al., 2004). To suppress background contributions, passive shields and active shields are usually set, including establishing graded shields of different materials such as Pb, Cu, Cd, polythene and so on, and adopting anti-veto methods such as anti-cosmic-ray systems and anti-Compton systems

*Corresponding author: Zeng Zhi, tel: +86 010 62796715; email address: zengzhi@tsinghua.edu.cn.

(Heusser, 1995; Kohler et al., 2009; Laurec et al., 1996; Zastawny, 2003). If possible, locating low background spectrometers underground is preferable while thick rock overburdens can almost eliminate the influence of cosmic rays.

To further reduce the background level, a low level HPGe gamma spectrometer, called GeTHU, was developed at China JinPing underground Laboratory (CJPL) in 2012, with a rock overburden of 2400 m (Cheng et al., 2011; Kang et al., 2010). The muon flux in CJPL is about $2.0 \times 10^{-6}$ m$^{-2}$s$^{-1}$ which is reduced by a factor of about $10^8$ compared to sea level (about 180 m$^{-2}$s$^{-1}$) (Wu et al., 2013). GeTHU is the first low level gamma spectrometer at CJPL and will be dedicated in material selections for CDEX dark matter detection experiment (Kang et al., 2013) and other rare event experiments as well as measurements of environmental samples with low radioactivity.

2. GeTHU facility

GeTHU is based on a coaxial n-type HPGe detector of 40% relative detection efficiency and was manufactured by CANBERRA in France. The germanium crystal has a height of 59.8 mm and a diameter of 59.9 mm. The cryostat is made of ultra-low background aluminum (ULB Al) to reduce the background and of U-shape to avoid direct line of sight from outside to the crystal. The preamplifier is positioned outside the shield due to the relatively high radioactivity. The energy resolution of the detector, defined as the ratio of the full width of half maximum (FWHM) to the energy of the gamma line, is shown in Figure 1. The FWHM of the detector at 1332.5keV is 2.04keV.

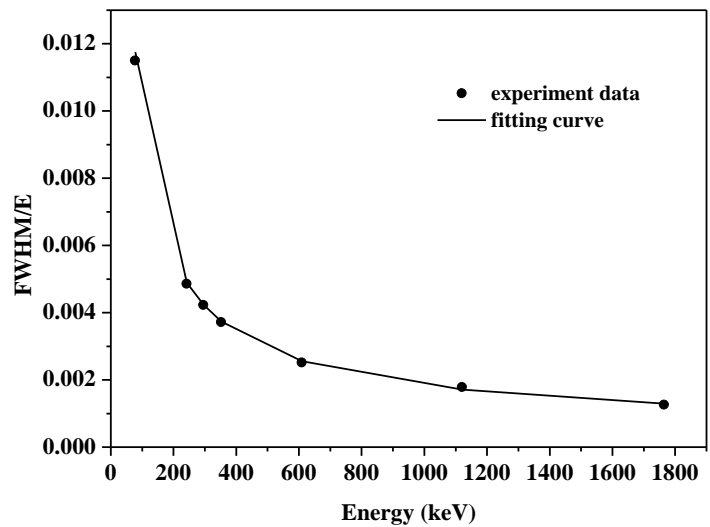

Figure 1 GeTHU's energy resolution. The solid curve is a fit with the function $\text{FWHM} = 0.5472 + 0.0408 \times \sqrt{E} + 4.218 \times 10^{-7} \times E^2$.

The passive graded shield of GeTHU, shown in Figure 2, has been custom-designed to offer a low background, a large sample capacity and easy access to the detector. The inner shield surrounding the measurement chamber is 5 cm of oxygen-free high purity copper (22 cm for the base plate), which has been polished with sandpaper and cleaned with anhydrous ethanol to remove residual surface contamination. Three layers of ordinary lead with a $^{210}$Pb activity of about 100 Bq/kg, each 5 cm thick, are surrounding the copper. All lead bricks were cleaned with anhydrous ethanol before being installed in the shield. The outermost is 10 cm of borated polyethylene plates which can prevent the penetration of ambient neutrons. The upper copper plate which supports the upper lead bricks and polyethylene plates is lain on sliding rails so that the sealed measurement chamber can be open and closed with the help of a hydraulic transmission equipment. The design of the shields paid special attention to avoiding direct line-of-sight to the crystal. The measurement chamber, with a dimension of $30 \times 30 \times 63 \mathrm{cm}^3$, has a capacity of large volume samples such as an 8 L Marinelli beaker. The entire facility is flushed with boil-off nitrogen from a dedicated LN$_2$ tank to reduce the influence from $^{222}$Rn in the air. With N$_2$ flushing, the $^{222}$Rn concentration in the air inside the measurement chamber can decrease from 100 Bq/m$^3$ to 1~4 Bq/m$^3$, resulting in an apparent background reduction in GeTHU spectra.

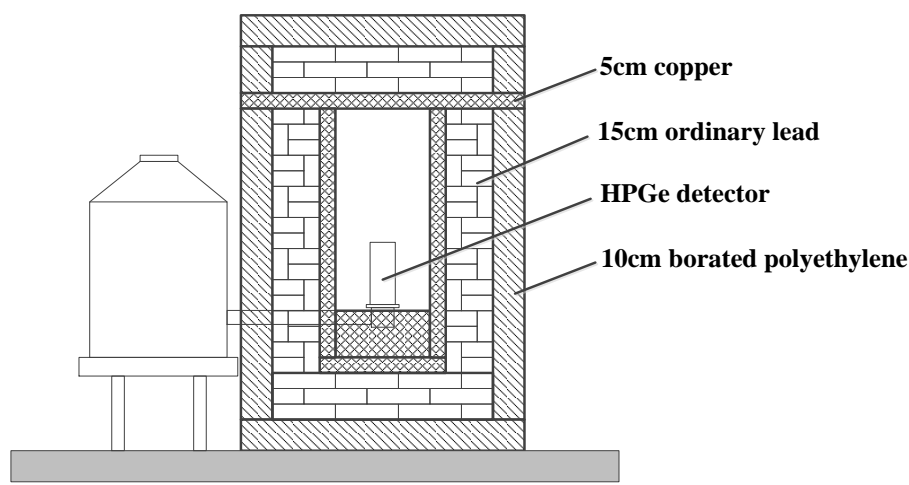

Figure 2   Schematic view of the GeTHU spectrometer at CJPL.

3. Results and discussion

3.1. Background characteristics

A 12-day background measurement of GeTHU was completed with nitrogen flushing in December 2012 and is shown in Figure 3. In the whole spectrum, all gamma peaks are induced by primordial radionuclides whereas no gamma peaks from artificial radionuclides are found, which proves no artificial radioactive contamination in construction materials. Some gamma peaks are prominent, such as the 46.5 keV peak of $^{210}$Pb mainly from the lead shield; the 295 keV, 352 keV ones of $^{214}$Pb and the 609 keV, 1765 keV ones of $^{214}$Bi from the shield materials or $^{222}$Rn in the air;

the 1461 keV one of $^{40}$K mainly from rocks and so on. No signs of cosmogenic radionuclides, such as $^{57}$Co, $^{58}$Co, $^{65}$Zn and so on (Kohler et al., 2009; Loaiza et al., 2011), are found in the spectrum, which is thanks to the long storage time of GeTHU in CJPL, about 2 years, during which most short living cosmogenic radionuclides have decayed. Table 1 gives the background count rates of gamma peaks from important primordial and artificial radionuclides as well as the integral background count rate between 40 and 2700 keV.

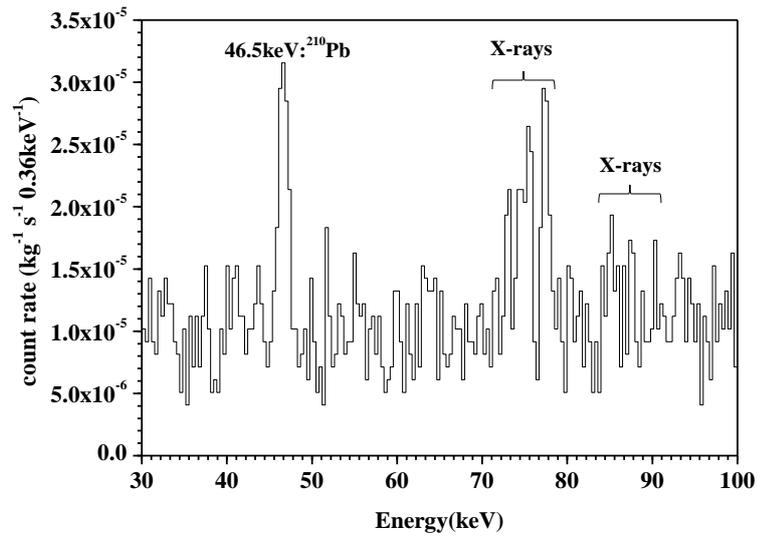

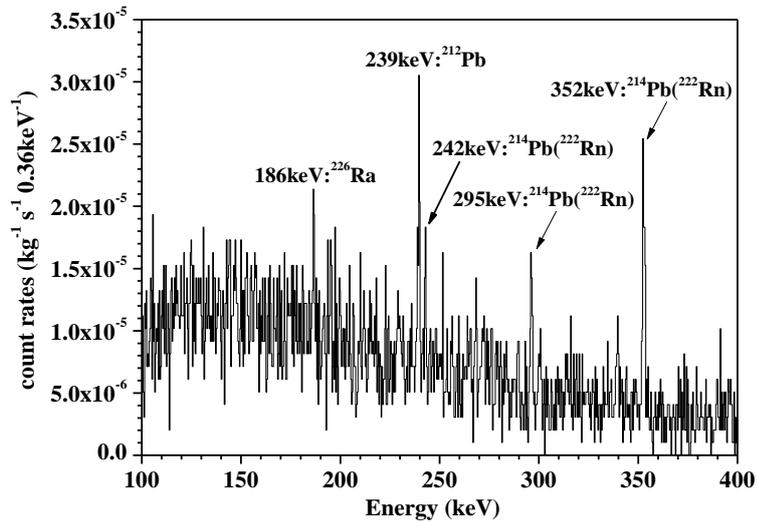

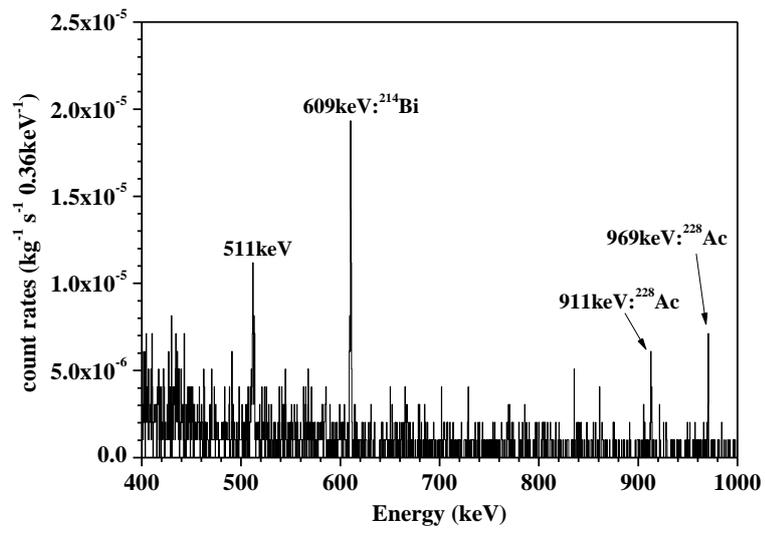

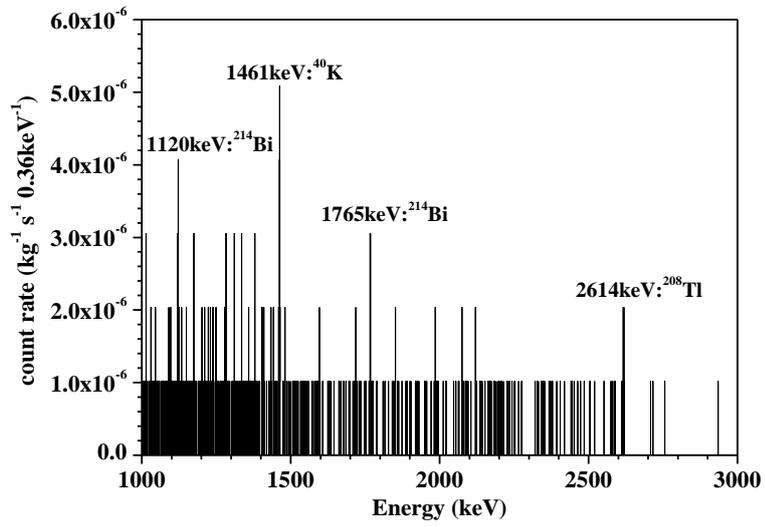

Figure 3 Background spectrum of GeTHU measured for 12 days at CJPL.

Table 1  Background count rates in the ±3σ regions of gamma peaks from main primordial and artificial radionuclides. Uncertainties are purely statistical.

| Energy (keV) | Nuclide | Peak/Integral count rate (day$^{-1}$) |
|---|---|---|
| **239** | $^{212}$Pb | 5.11±1.20 |
| **352** | $^{214}$Pb | 6.09±0.96 |
| **511** | $\beta^+$ | 3.85±0.71 |
| **609** | $^{214}$Bi | 5.10±0.73 |
| **662** | $^{137}$Cs | <1.49 |
| **911** | $^{228}$Ac | <1.86 |
| **1120** | $^{214}$Bi | <1.73 |
| **1173** | $^{60}$Co | <1.54 |
| **1332** | $^{60}$Co | <1.26 |
| **1461** | $^{40}$K | <1.90 |
| **1765** | $^{214}$Bi | <1.43 |
| **40~2700** | | 815.6 |

In order to understand the origin of the remaining background, a simulated background spectrum was obtained with Monte Carlo methods, in which gamma rays resulting from the $^{238}$U series, $^{232}$Th series and $^{40}$K in the lead and copper shields were taken into consideration as well as beta rays from $^{210}$Bi, the daughter of $^{210}$Pb, in the lead shields and from natural radioactive series in the copper shields, due to the possible influence of bremsstrahlung production. The simulated result is shown in Figure 4 and compared to the 12-day spectrum. It is clear that in the measured background spectrum, the 239 keV gamma peak is higher than the 352 keV one, which is again higher than the 609 one, while the simulated background spectrum shows that the 609 keV gamma peak is a little higher than the 352 keV one. The differences indicate that the background from the $^{238}$U series, $^{232}$Th series and $^{40}$K in the shields is not the dominant part and taking into account that the 352 and 609 keV gamma peaks both come from the daughters of $^{222}$Rn, it can be concluded that the $^{222}$Rn in the air around the detector must contribute significantly to the background.

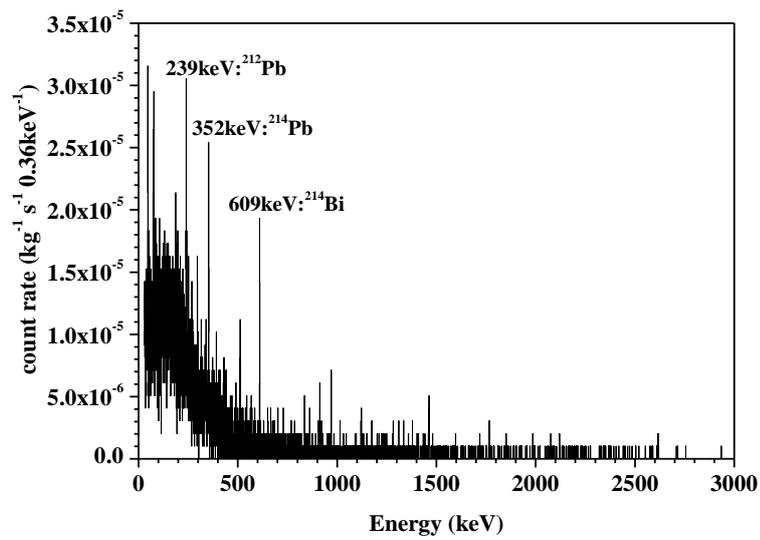

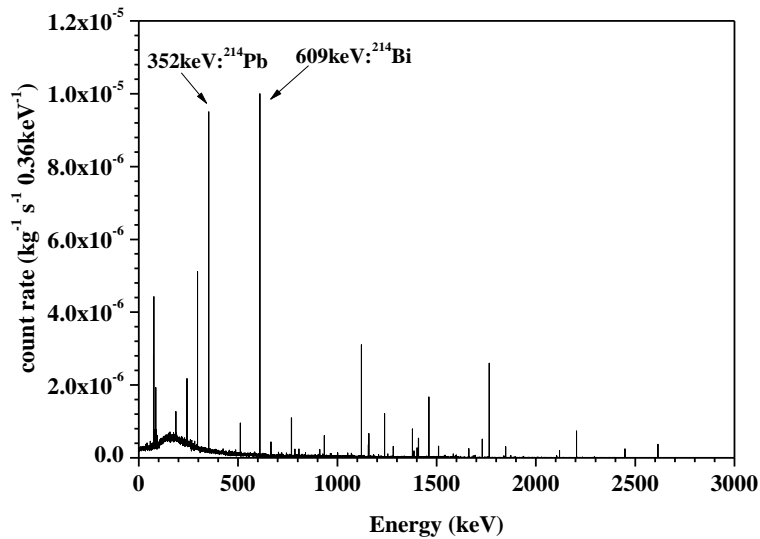

Figure 4  Comparison of the measured background spectrum of GeTHU (top) with the one obtained from Monte Carlo simulation resulting from primordial nuclides in copper and lead (bottom).

To further understand the influence of $^{222}$Rn, measurements of background without nitrogen flushing were carried on as well as the measurement with nitrogen flushing and results are shown in Figure 5. It is obvious that, after flushing with nitrogen, the background level dropped significantly due to the reduction of $^{222}$Rn concentration in the measurement chamber and the integral count rate between 100~2700 keV was only 4% of that without nitrogen flushing. Other measures will be developed to further reduce the background from $^{222}$Rn in the future.

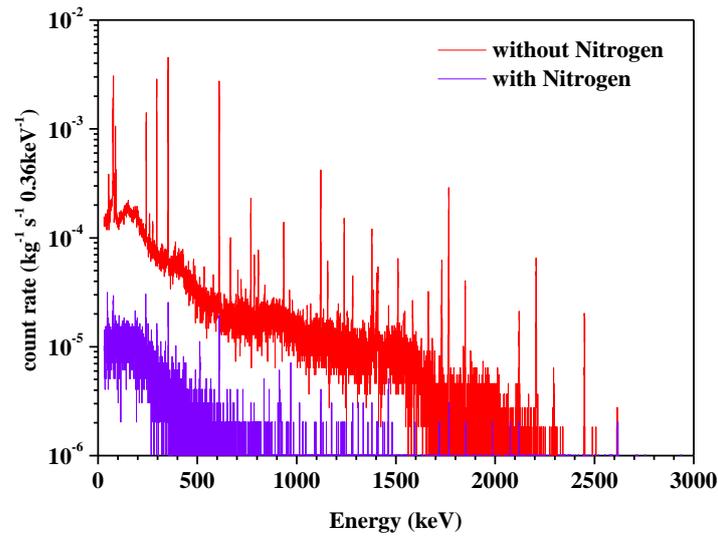

Figure 5   Background spectra of GeTHU with (blue) and without (red) N2 flushing at CJPL.

When measuring the 12-day background spectrum with nitrogen flushing, the activity concentration of $^{222}$Rn in the experiment hall rose markedly because of the malfunction of the ventilation system. Especially in the first 4 days the concentration was up to about 300 Bq/m$^3$, while in the last 8 days the concentration was only about 80 Bq/m$^3$. Accordingly, the background spectrum only for the last 8 days was separated from the whole one to study the impact of the high radon concentration in the experiment hall. The integral count rates in the peak regions of main gamma rays induced by $^{222}$Rn, obtained both from the whole spectrum and the 8-day one, are shown in Table 2. Apparently, corresponding integral count rates of interesting gamma peaks obtained from both two spectrums are close to each other and there is no obvious fluctuation in the background level. Therefore, it is effective to flush the measurement chamber with nitrogen, which can keep the $^{222}$Rn concentration inside immune to the change of outside environment and stable at a low level.

Table 2  Comparison of the count rates of single gamma rays induced by $^{222}$Rn with normal/fault condition of ventilation system in CJPL.

| Energy (keV) | Nuclide | Peak count rate (day$^{-1}$) | |
| --- | --- | --- | --- |
| | | 12-day (from 12/12 to 24/12) | 8-day (from 16/12 to 24/12) |
| 242 | $^{214}$Pb | <3.55 | <4.45 |
| 295 | $^{214}$Pb | <3.36 | <4.04 |
| 352 | $^{214}$Pb | 6.09±0.96 | 6.95±1.19 |
| 609 | $^{214}$Bi | 5.10±0.73 | 6.30±0.91 |
| 1120 | $^{214}$Bi | <1.73 | <2.26 |
| 1765 | $^{214}$Bi | <1.43 | <1.93 |

3.2. Detection limits

Detection limits for selected peak energies of $^{214}$Pb, $^{214}$Bi, $^{228}$Ac, $^{212}$Pb, $^{212}$Bi, $^{40}$K, $^{137}$Cs and $^{60}$Co are presented here as evidences of the facility's performance to detect them. The detection limit, $L_d$, was calculated by the following equation with a 95% confidence level (Gilmore et al., 1995):

$$L_d = 2.71 + 4.65 \times \sqrt{B} \qquad [1]$$

where B represents the background counts in a region of 2.54 FWHM around a specific peak of interest. Results are given in Table 3, where $L_d$ was calculated from the 12-day background spectrum.

Table 3  Results of $L_d$.

| Nuclide | Energy (keV) | $L_d$ (counts) |
| --- | --- | --- |
| $^{212}$Pb | 239 | 54.49 |
| $^{214}$Pb | 352 | 51.92 |
| $^{214}$Bi | 609 | 42.71 |
| $^{137}$Cs | 662 | 18.82 |
| $^{212}$Bi | 727 | 15.86 |
| $^{228}$Ac | 911 | 23.51 |
| $^{60}$Co | 1173 | 19.48 |
| $^{60}$Co | 1332 | 15.86 |
| $^{40}$K | 1461 | 24.02 |
| $^{214}$Bi | 1765 | 18.13 |

Table 3 shows a good performance of GeTHU to detect the selected 8 radionuclides, most of whose detection limits are in the magnitude of some tens of counts. In general, the detection limit goes down as the energy goes up due to the higher background level in the low energy region than in the high region.

3.3. Efficiency calibration

In general measurements of samples, efficiencies are of great importance in the analysis of interesting radionuclides. However, given the various geometries and components of samples, making reference samples for each measurement is obviously not feasible and always difficult in efficiency calibrations. Consequently, other methods are applied instead, among which Monte Carlo simulation method is a good alternative.

With simulations, one can either obtain efficiencies of point sources first and then use analytical methods to calculate that of complex geometries, or obtain the final efficiencies directly (Chen et al., 2013). This is often called "sourceless efficiency calibrations" on most occasions. With the formula below, the efficiencies can be calculated:

$$\varepsilon = \frac{A_{net}}{N} \quad [2]$$

where ε stands for the efficiency of a specific energy, $A_{net}$ represents the net counts of the corresponding full energy peak and N correspond to the number of simulated events.

Figure 6 shows a schematic view of the detector of GeTHU. In this preliminary work, point sources are employed in the simulation of efficiencies and finally the energy dependency of efficiencies is shown with a curve (see Figure 7). In simulations, point sources are located 25 cm above the endcap assuming isotropic emissions and a series of gamma rays from $^{238}$U series, $^{232}$Th series, $^{40}$K and some artificial radionuclides such as $^{152}$Eu is used. In Figure 7, the curve is fitted with the following function (Guan and Zhou, 2006):

$$\ln \varepsilon = a_1 \times E + a_2 + a_3 \times \ln E/E + a_4/E + a_5/E^2 + a_6/E^3 + a_7 \times \ln E \quad [3]$$

where ε represents efficiencies and E stands for the energy of gamma rays. The efficiency of 1.33 MeV gamma ray from $^{60}$Co is calculated using the final fitting function and transformed into the relative efficiency, which is about 0.45 while the standard data from the manufacture is 0.4. The main cause of this error between two data is the inaccuracy of the detector model in the simulation process. Further research will be implemented on the efficiency simulation of complex geometries and corresponding measurements will be done at the same time.

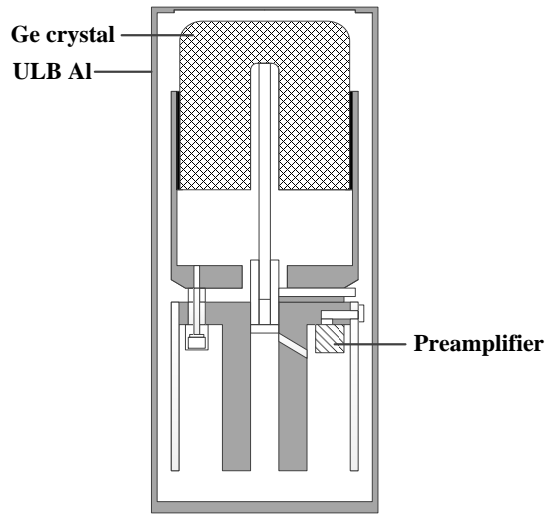

Figure 6　Schematic view of the detector of GeTHU.

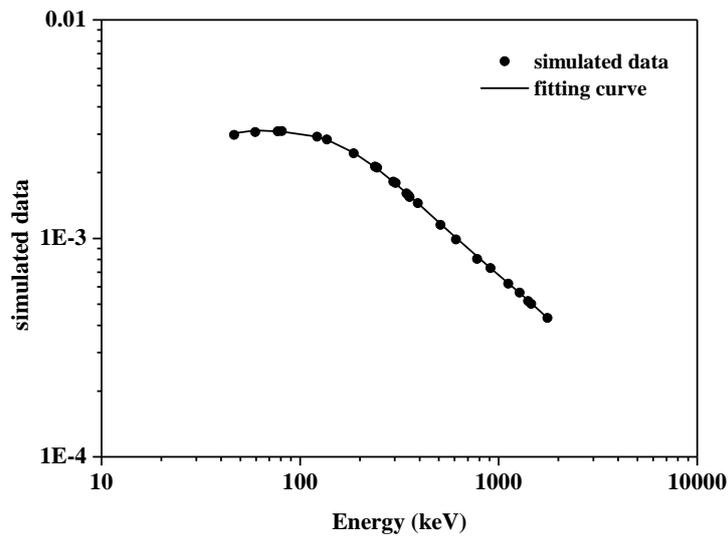

Figure 7　The efficiency as a function of energy. The solid line represents a fit using equation [3].

### 3.4. Sample measurements

Boric acid and silica sand are both crucial raw materials in today's industrial production, while the former can act as disinfectants, pesticides, be used to produce glass, leather or even absorbing rods in nuclear power plants, and the latter usually plays important roles in metallurgy, construction, electronics and other areas. Thus it is critical to investigate the radioactivity in such materials and ensure that no unexpected contamination occurs.

With GeTHU, collected boric acid and silica sand samples were measured for about 3.2 days and 2 days respectively and the spectra are shown in Figure 8. Specific radioactivity for important primordial and artificial radionuclides was calculated with the function that

$$A_m = \frac{N_{net}}{\varepsilon \cdot t \cdot \gamma \cdot m} \quad [4]$$

where $A_m$ represents specific radioactivity, $N_{net}$ stands for net counts of the selected full energy peak, $\varepsilon$ is the full energy detection efficiency obtained with Monte Carlo methods, t is the measurement time, $\gamma$ is the emission probability and m is the sample mass. Results of radioactivity are given in Table 4. It shows that the radioactivity of $^{238}$U and $^{232}$Th series in boric acid is higher than that in silica sand, which indicates the radio-impurity of the boric acid. What's more, there is some $^{137}$Cs contamination in the boric acid which must cause sufficient attention when selecting boric acid as raw materials.

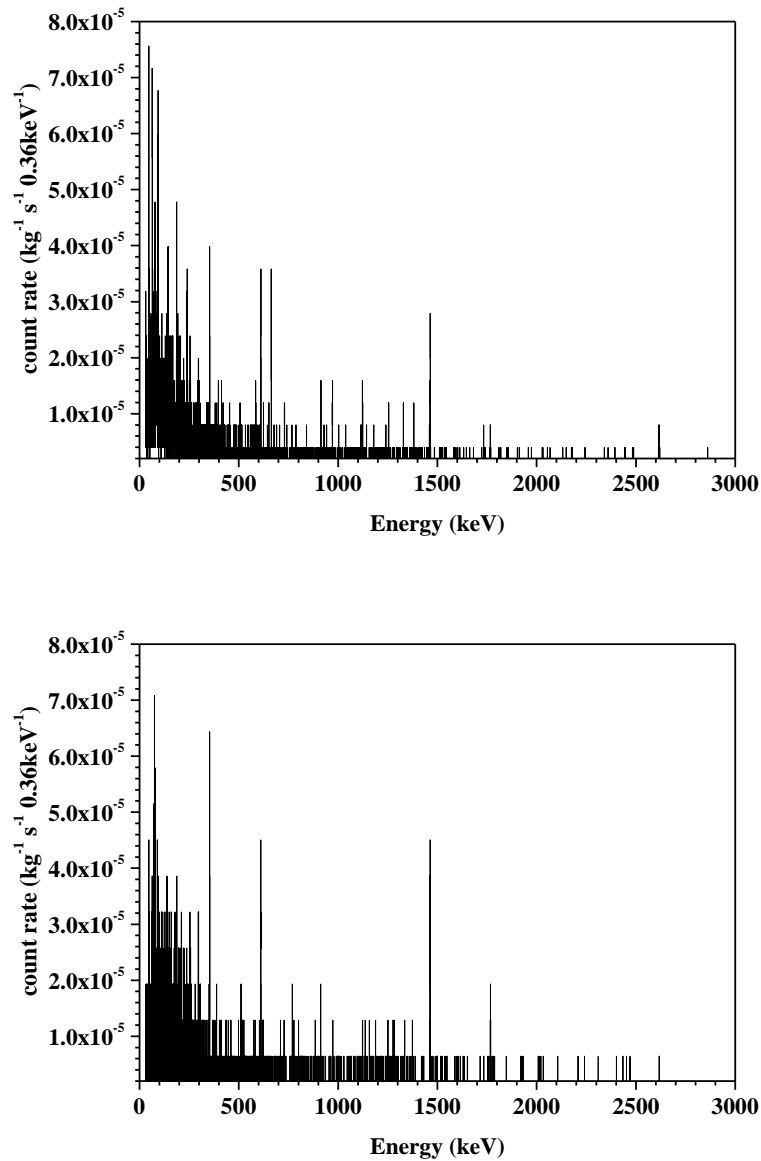

Figure 8    Measured spectra for boric acid (top) and silica sand (bottom).

Table 4  Radioactivity of some radionuclides in the boric acid and silica sand samples. Uncertainties are purely statistical.

| Series | Nuclide | Energy (keV) | Activity (mBq/kg) | |
|---|---|---|---|---|
| | | | Boric acid | Silica sand |
| $^{238}$U | $^{234}$Th | 63.29 | 258±61 | 121±54 |
| $^{238}$U | $^{234m}$Pa | 1001.03 | 414±337 | <772 |
| $^{238}$U | $^{226}$Ra | 186.221 | <209 | <151 |
| $^{238}$U | $^{214}$Pb | 295.21 | <30 | <35 |
| $^{238}$U | $^{214}$Pb | 351.92 | <21 | <24 |
| $^{232}$Th | $^{212}$Pb | 238.632 | 17±8 | <10 |
| $^{232}$Th | $^{228}$Ac | 911.07 | 34±16 | 26±12 |
| $^{232}$Th | $^{208}$Tl | 583.14 | 11±4 | <6 |
| | $^{40}$K | 1460.8 | 311±111 | 304±103 |
| | $^{137}$Cs | 661.65 | 21±6 | <3 |

4. Conclusion

A new low background germanium spectrometer, GeTHU, has been installed and run at CJPL. The integral background count rate from 40 to 2700 keV is about 0.6 cpm and the detection limits for the main gamma rays of some primordial and artificial radionuclides are in the level of some tens of counts based on a 12-day background spectrum. $^{222}$Rn in the air around the detector contributes significantly to the remaining background and attention will be paid to the radon concentration especially inside the sample chamber. Some material samples have been screened with GeTHU and $^{137}$Cs contamination was found in the boric acid. GeTHU will be used to measure environmental samples and serve for material selections of CDEX experiment (Kang et al., 2013).

Acknowledgements

This work is partly supported by National Science Foundation of China through No.11175099, 11075090 and 11055002. We are grateful to Mr. Qin Jianqiang, He Qingju and other colleagues of CDEX collaboration for their help during the design and installation of the spectrometer.

References

Arpesella, C., Back, H., Balata, M., Beau, T., Bellini, G., Benziger, J., Bonetti, S., Brigatti, A., Buck, C., Caccianiga, B., 2002. Measurements of extremely low radioactivity levels in BOREXINO. Astroparticle Physics 18, 1-25.

Budjáš, D., Gangapshev, A., Gasparro, J., Hampel, W., Heisel, M., Heusser, G., Hult, M., Klimenko, A., Kuzminov, V., Laubenstein, M., 2009. Gamma-ray spectrometry of ultra low levels of radioactivity within the material screening program for the GERDA experiment. Applied Radiation and Isotopes 67, 755-758.

Chen, L., Ma, H., Zeng, Z., Li, J.L., Cheng, J.P., 2013. Monte Carlo-based sourceless efficiency calibration method of HPGe γspectrometer. High Power and Paeticle beams 25, 201.

Cheng, J.P., Wu, S.Y., Yue, Q., Shen, M.B., 2011. A review of international underground laboratory developm ents. Physics 40, 149-154.

Gilmore, G., Hemingway, J.D., Gilmore, G., 1995. Practical gamma-ray spectrometry. Wiley Chichester.

Guan, Y.X., Zhou, G., 2006. The Semi-empirical Formula Fitting of Germanium Detector Efficiency Curve. Journal of Zaozhuang University 23, 77-79.

Heusser, G., 1995. Low-radioactivity background techniques. Annual Review of Nuclear and Particle Science 45, 543-590.

Hult, M., 2007. Low-level gamma-ray spectrometry using Ge-detectors. Metrologia 44, S87-S94.

Hult, M., Preusse, W., Gasparro, J., Kohler, M., 2006. Underground gamma-ray spectrometry. Acta Chimica Slovenica 53, 1.

Kang, K.J., Cheng, J.P., Chen, Y.H., Li, Y.J., Shen, M.B., Wu, S.Y., Yue, Q., 2010. Status and prospects of a deep underground laboratory in China. Journal of Physics: Conference Series 203, 012028.

Kang, K.J., Cheng, J.P., Li, J., Li, Y.J., Yue, Q., Bai, Y., Bi, Y., Cheng, J.P., Chen, N., Chen, N., 2013. Introduction of the CDEX experiment. arXiv preprint arXiv:1303.0601.

Kohler, M., Degering, D., Laubenstein, M., Quirin, P., Lampert, M.O., Hult, M., Arnold, D., Neumaier, S., Reyss, J.L., 2009. A new low-level gamma-ray spectrometry system for environmental radioactivity at the underground laboratory Felsenkeller. Applied radiation and isotopes : including data, instrumentation and methods for use in agriculture, industry and medicine 67, 736-740.

Laubenstein, M., Hult, M., Gasparro, J., Arnold, D., Neumaier, S., Heusser, G., Köhler, M., Povinec, P., Reyss, J.-L., Schwaiger, M., 2004. Underground measurements of radioactivity. Applied radiation and isotopes 61, 167-172.

Laurec, J., Blanchard, X., Pointurier, F., Adam, A., 1996. A new low background gamma spectrometer equipped with an anti-cosmic device. Nuclear Instruments and Methods in Physics Research Section A: Accelerators, Spectrometers, Detectors and Associated Equipment 369, 566-571.

Loaiza, P., Chassaing, C., Hubert, P., Nachab, A., Perrot, F., Reyss, J.L., Warot, G., 2011. Low background germanium planar detector for gamma-ray spectrometry. Nuclear Instruments and Methods in Physics Research Section A: Accelerators, Spectrometers, Detectors and Associated Equipment 634,

**Figures**

Figure 1   GeTHU's energy resolution. The solid curve is a fit with the function $FWHM = 0.5472 + 0.0408 \times \sqrt{E + 4.218 \times 10^{-7} \times E^2}$.

Figure 2   Schematic view of the GeTHU spectrometer at CJPL.

Figure 3   Background spectrum of GeTHU measured for 12 days at CJPL.

Figure 4   Comparison of the measured background spectrum of GeTHU (top) with the one obtained from Monte Carlo simulation resulting from primordial nuclides in copper and lead (bottom).

Figure 5   Background spectra of GeTHU with (blue) and without (red) N2 flushing at CJPL.

Figure 6   Schematic view of the detector of GeTHU.

Figure 7   The efficiency as a function of energy. The solid line represents a fit using equation [3].

Figure 8   Measured spectra for boric acid (top) and silica sand (bottom).

**Tables**

Table 1   Background count rates in the ±3σ regions of gamma peaks from main primordial and artificial radionuclides. Uncertainties are purely statistical.

Table 2   Comparison of the count rates of single gamma rays induced by $^{222}$Rn with normal/fault condition of ventilation system in CJPL.

Table 3   Results of $L_d$.

Table 4   Radioactivity of some radionuclides in the boric acid and silica sand samples. Uncertainties are purely statistical.